\documentclass[prl,aps,twocolumn,nosuperscriptaddress,bibnotes,notitlepage,nofootinbib]{revtex4-1}

\usepackage[utf8]{inputenc}
\usepackage{amsmath}
\usepackage{float}
\usepackage{subfigure}
\usepackage{color}
\usepackage{natbib}
\usepackage{graphicx}
\usepackage{caption}
\usepackage[normalem]{ulem}
\newcommand{\magg}{\mathbf{M}}
\newcommand{\stagg}{\mathbf{N}}
\newcommand{\spin}{\mathbf{S}}

\newcommand{\abc}{\alpha \beta \gamma}
\newcommand{\comnt}[1]{\ignorespaces}

\usepackage[colorlinks=true,linkcolor = blue,
            urlcolor  = blue,
            citecolor = blue,
            anchorcolor = blue]{hyperref}

\begin{document}

\title{Emergent hydrodynamics in a non-reciprocal classical isotropic magnet}

\author{Nisarg Bhatt}
\email{bhatt@iisc.ac.in}
\author{Subroto Mukerjee}
\email{smukerjee@iisc.ac.in}
\author{Sriram Ramaswamy}
\email{sriram@iisc.ac.in}
\affiliation{Centre for Condensed Matter Theory, Department of Physics, Indian Institute of Science, Bangalore 560 012, India}

\begin{abstract}
The Hamiltonian nature of the precessional dynamics of the classical Heisenberg model leads to reciprocal interactions amongst the spins. Heisenberg spins is reciprocal in nature. 
In this work, we study the dynamics of a nonequilibrium classical 
spin chain in which 
the neighbours interact through a purely non-reciprocal exchange coupling [\href{https://iopscience.iop.org/article/10.1209/epl/i2002-00280-2}{EPL \textbf{60}, 418 (2002)}] which preserves rotational symmetry. The resultant dynamics conserves neither magnetization nor energy. We uncover other local conservation laws in their place in the extreme case of a strictly antisymmetric coupling. We show
numerically that the model undergoes an analogue of thermalization. We present results on the presence of conserved quantities, their diffusive spreading and a hydrodynamic picture, and the nature of the decorrelation front upon adding an initial perturbation to the system.

\end{abstract}
\maketitle

\paragraph{Introduction:} Effective interactions that evade Newton's 3rd Law are ubiquitous in living or driven matter \cite{das2002driven,ivlev2015statistical,dadhichi2020nonmutual,SSaha_etalPRX2020,you2020nonreciprocity,CahnHilliardnonvarcoupling2021,PhysRevLett.104.178103}.
The importance of such non-reciprocity (hereafter NR) has long been appreciated in the context of learning and associative memory \cite{parisi1986asymmetric, PhysRevE.86.011909}, and the subject is currently experiencing a resurgence of interest \cite{fruchart2021non, bowick2022symmetry, shamikgupta2019}.

The breaking of reciprocity leads to a host of highly original effects including oscillations and travelling waves without inertia, advection without a velocity, pursuit-and-capture behaviour, and exceptional-point phase transitions 
\cite{ PhysRevLett.106.054102, PhysRevLett.117.248001, PhysRevE.99.012612, PhysRevLett.122.185301, PhysRevResearch.2.033018, Galda2019}. 
An early study \cite{das2002driven} explored criticality, chaos and control in the reversible precessional dynamics of a lattice of classical Heisenberg spins, with distinct exchange couplings for right and left neighbours. It was argued \cite{das2002driven} that this manifestly non-reciprocal and non-Hamiltonian dynamics should arise generically in the presence of a sustained energy throughput, if the underlying lattice was non-centrosymmetric. 

The \textit{canonical} dynamics of classical Heisenberg spins \cite{ma1975critical} $\{\spin_i\}$ on a lattice  consists of Larmor precession: $d\spin_i/dt = \gamma \spin_i \times \mathbf{h}_i$, 
where $\gamma$ is the gyromagnetic ratio and the local magnetic field $\mathbf{h}_i = \partial H / \partial \spin_i$ arises from interactions with neighboring spins through a Hamiltonian $H = - \sum_{i,j}J_{ij}\spin_i \cdot \spin_j$. The reciprocal nature of this dynamics is clear: the torque on spin $i$ due to spin $j$ is precisely the negative of that on $j$ due to $i$. The model of \cite{das2002driven} amounts to retaining precessional dynamics, with $\mathbf{h}_i = \sum_j J_{ij} \spin_j$, but allowing a non-symmetric $J_{ij}$. The dynamics still preserves spin-rotational symmetry but cannot be derived from a Hamiltonian.

In this Letter, we study a chain of Heisenberg spins with purely antisymmetric couplings, $J_{ij} = - J_{ji}$. Despite the absence of a Hamiltonian description for this maximally non-reciprocal dynamics, we find the system displays 
an analogue of thermalization and smooth fluctuating hydrodynamic behaviour of emergent conserved fields, namely the three components of the \textit{staggered} magnetization and a scalar which we call the pseudo-energy. These are distinct from quasiconserved quantities as discovered in a class of periodically driven systems \cite{McRoberts_PRR5}. We develop a self-consistent treatment which leads to diffusion for these slow fields and relaxational dynamics for the regular magnetization, which we corroborate by direct numerical simulation of the microscopic equations of motion. Finally, we numerically calculate the decorrelator for our system and show that it displays a ballistic spreading of chaos like its Hamiltonian counterpart thereby providing further evidence for thermalization.

The decorrelator or Out-of-Time-Ordereded-Correlator (OTOC) has proven to be a useful tool in the study of chaos spreading in quantum non-integrable systems and also classical ones with elementary degrees of freedom, like spins, that do not have a symplectic Poisson bracket structure.   \cite{PhysRevB.105.214308,PhysRevX.8.031057,1503.01409}. The decorrelator of a classical one-dimensional system of Heisenberg spins described by a Hamiltonian with only nearest neighbour exchange (henceforth referred to as the Heisenberg model) was found to exhibit a ballistic spreading of chaos with a butterfly velocity even while the magnetization and energy densities displayed diffusion at high temperature, consistent with the expectation of thermalization in this system~\cite{das2018light,McRoberts_PRB105}.

\begin{figure*}[htp]
\centering
\subfigure[$C_{N}(x,t)$ ]
  {\includegraphics[scale=0.28]{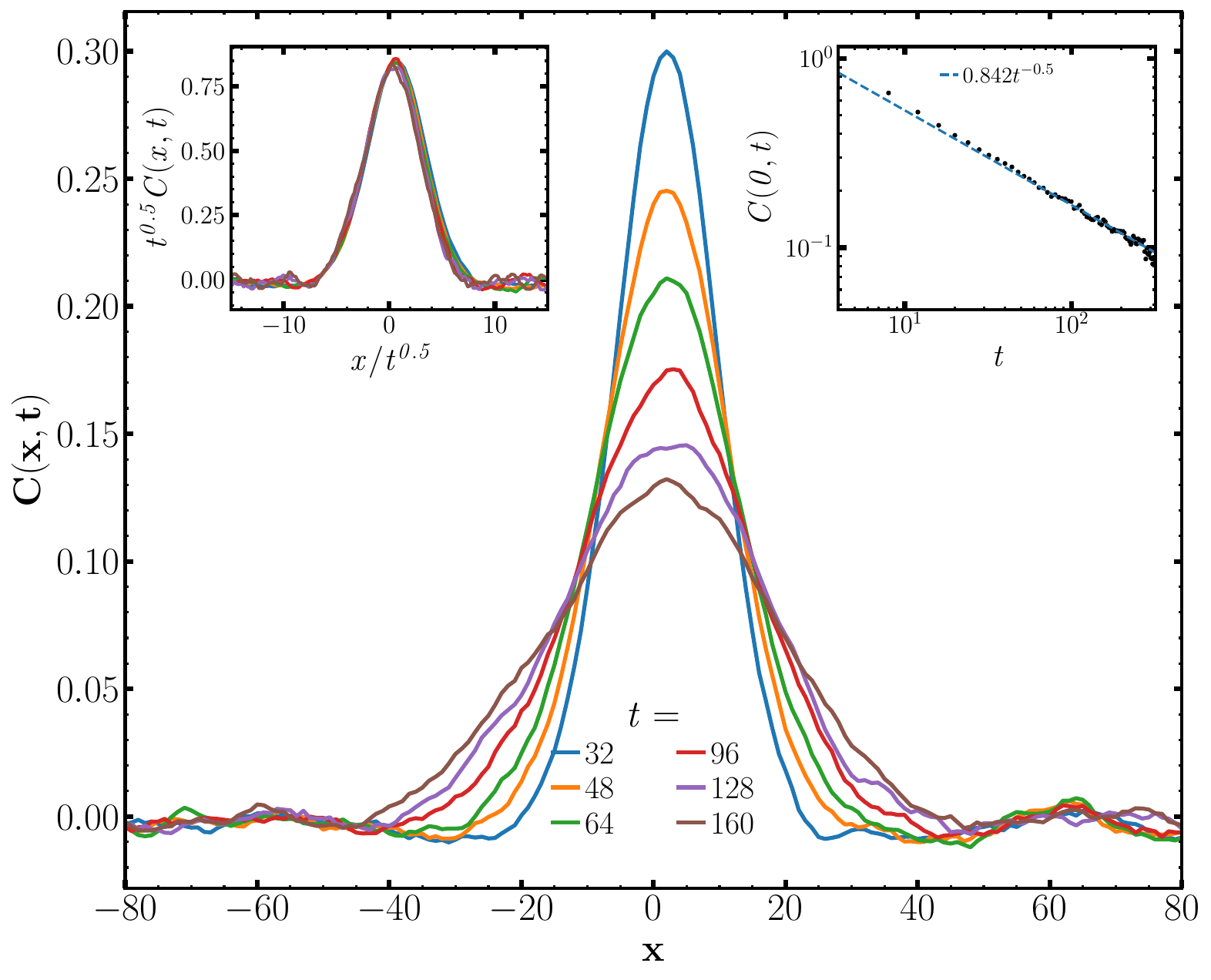}}
  \label{fig:Nxt_correlations}
\quad \quad \quad 
\subfigure[$C_{{\mathcal{E}}}(x,t)$]  
    {\includegraphics[scale=0.28]{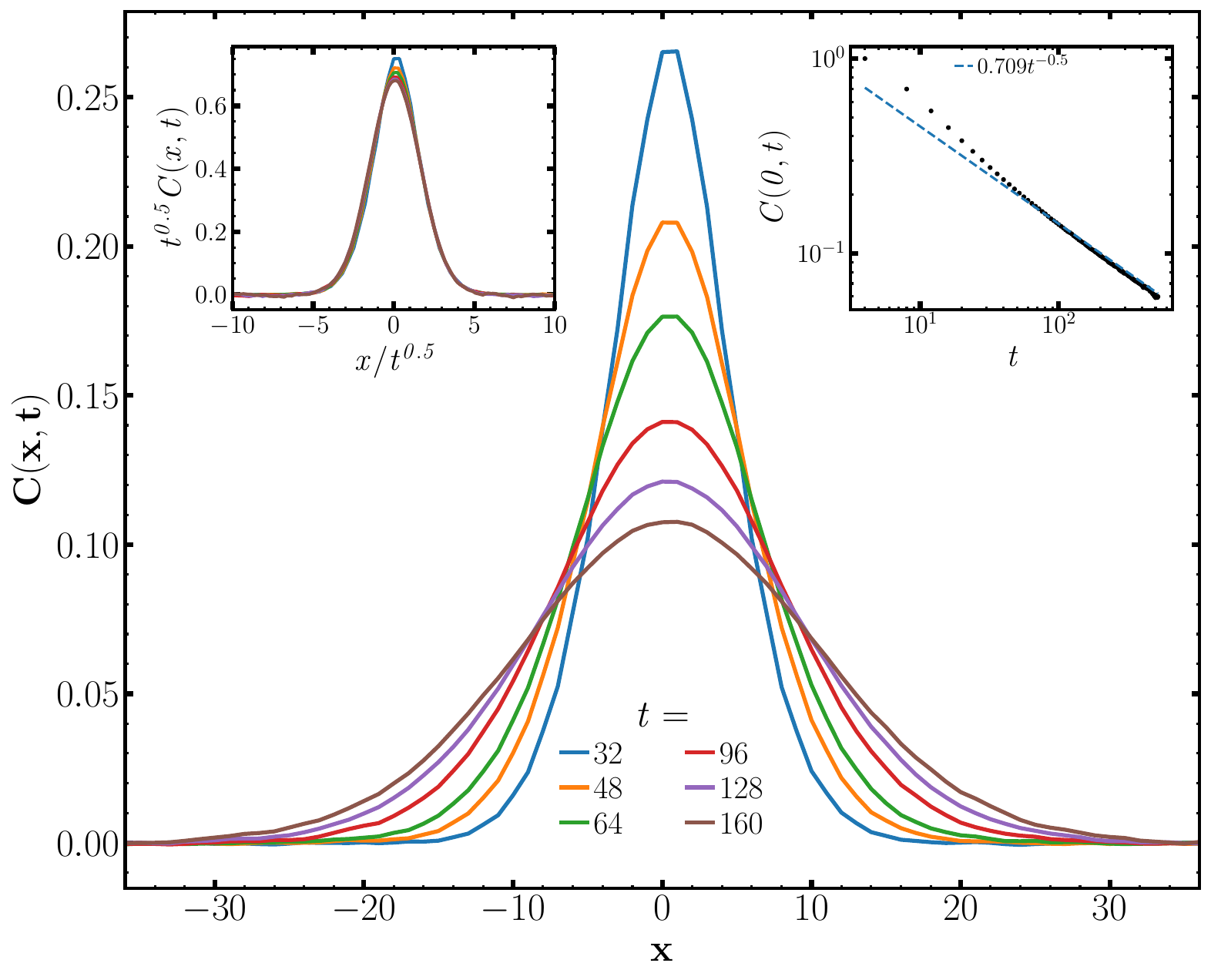}}
     \label{fig:CENxt_correlations}
\caption{ \small Two-point correlation functions of: (a) the staggered magnetization, $C_{N}(x,t)$ and (b) the pseudo-energy , $C_{\mathcal{E}}(x,t)$, with $x \in \{1,L\}$ for $L=512$, sampled over 5000 initial configurations as functions of $x$ for different values of $t$. The left inset of both panels shows the scaling collapse to a form $C(x,t) = t^{-1/2}f(x/t^{1/2})$ consistent with diffusion while the right shows a plot of $C(0,t)$ versus $t$ with a fit to $t^{-1/2}$.}
\label{fig:NN_EN_corrxt}
\end{figure*}

\paragraph{Model:} We study a system of classical spins ${\spin}_i$ of unit magnitude on the sites $i$ of a one-dimensional lattice of $L$ sites with periodic boundary conditions. The equation of motion of the spins is 
\begin{align}
\dot{\spin_i} &= \spin_i \times (\spin_{i+1} - \spin_{i-1} )
\label{eqn:drvn_dynmk}
\end{align}
Although the dynamics is non-Hamiltonian, it  obeys a Liouville theorem \cite{Suppinf}, i.e., the flow in spin-space is divergence-free.
Eqn.(~\ref{eqn:drvn_dynmk}) conserves neither the magnetization ${\magg}=\sum_i {\spin}_i$ nor the Heisenberg Hamiltonian.
However, the dynamics conserves the staggered magnetization ${\stagg}=\sum_i (-1)^i {\spin}_i$ and the quantity 
$
\mathcal{E}= \sum_i (-1)^i {\spin}_i \cdot {\spin}_{i+1}$, which we refer to as the pseudo-energy. These are analogues of the magnetization and energy that are conserved in the Heisenberg model.
We numerically calculate the two-point correlators $C_N(x,t) = \langle {\stagg}(x+x_0,t+t_0) \cdot {\stagg}(x_0,t_0)\rangle$ and $C_{{\mathcal E}}(x,t) = \langle {\mathcal E}(x+x_0,t+t_0){\mathcal E} (x_0,t_0)\rangle$ in the statistical steady state, with the samples drawn from an ensemble with equally weighted microstates. 
The results are shown in Fig. (~\ref{fig:NN_EN_corrxt}) where the good data collapse shows that correlations of ${\stagg}$ and of ${\mathcal E}$ spread diffusively.
 
\paragraph{Hydrodynamic description:}To arrive at the continuity equations for the two conserved quantities, we rewrite the spins at the even and odd lattice indices as linear combinations of magnetization and staggered magnetization: 
$ \spin_{2i}  = \magg_i + \stagg_i ,  \spin_{2i+1} = \magg_i - \stagg_i$. 
The pseudo-energy can be similarly rewritten as 
${\mathcal{E}}_i = \frac{1}{2}\spin_{2i} \cdot (\spin_{2i-1} - \spin_{2i+1})
     = -\frac{1}{2}(\magg_i+\stagg_i)\cdot (\magg_{i-1} -  \stagg_{i-1} - \magg_i + \stagg_i) $.
The spin dynamics (~\ref{eqn:drvn_dynmk}) can be written in terms of $\magg_{i}$ and $\stagg_{i}$ and coarse-grained while retaining only the most relevant lower order gradient terms. In a na\"{\i}ve continuum limit, in which coarse-grained expressions for products of variables are replaced by the products of the individual coarse-grained quantities, we obtain the following hydrodynamic equations for $\magg$ and $\stagg$~\cite{Suppinf}.
\begin{align}
    \partial_t \magg(x,t) &=  \magg \times \partial_x \magg  - \stagg \times \partial_x \stagg
    \label{eqn:M_contnus}
\end{align}
\begin{align}
    \partial_t \stagg(x,t) &= \partial_x (\stagg \times \magg )
    \label{eqn:N_contnus}
\end{align}
Further, the staggered energy in the continuum limit is
${\mathcal{E}}(x,t)   =  -\frac{1}{2} (\magg + \stagg) \cdot \partial_x (\magg - \stagg) $, with the dynamical equation~\cite{Suppinf}
 \begin{align}
    \partial_t {\mathcal{E}} &= \partial_x \left( (\magg \times \stagg) \cdot \partial_x (\magg + \stagg) \right)
    \label{eqn:EN_contnus}
\end{align}
Eqns. (~\ref{eqn:N_contnus}) and (~\ref{eqn:EN_contnus}) have the form of continuity equations, as expected, for the conserved quantities $\stagg$ and ${\mathcal{E}}$, with the currents $j_{\stagg} = \magg \times \stagg $ and $j_{{\mathcal{E}}} =  -(\magg \times \stagg) \cdot \partial_x (\magg + \stagg)$ respectively. 

Equations (~\ref{eqn:M_contnus} - ~\ref{eqn:EN_contnus}) have been obtained for the hydrodynamic (i.e. low wavenumber) modes of $\magg$ and $\stagg$. The conservation law for $\stagg$ suggests that it is a slow variable evolving against the backdrop of the decaying non-conserved variable $\magg$. We expect that the effect of the large wavenumber (fast) modes on these hydrodynamic modes is to provide a noise for each, which in conjunction with the non-linearities in (~\ref{eqn:M_contnus}),(~\ref{eqn:N_contnus}) give rise to diffusion for the conserved mode $\stagg$ and relaxation for the non-conserved mode $\magg$.
We thus write down the following effective hydrodynamic equations for $\magg$ and $\stagg$.

\begin{align}
    \begin{split}
         \partial_t \magg &= -\magg/\tau + u_M (\magg \times \partial_x \magg) - u_N (\stagg \times \partial_x \stagg) + \boldsymbol{\xi} 
         \end{split}
         \label{eqn:Mhydro}
         \end{align}
and       
\begin{align}
    \begin{split}
             \partial_t \stagg &= D\partial_x^2 \stagg - u_{MN} \partial_x (\magg \times \stagg) + \boldsymbol{\zeta},
    \end{split}
    \label{eqn:Nhydro}
\end{align}
where $\boldsymbol{\xi}$ and $\boldsymbol{\zeta}$ are a non-conserving and conserving noise consistent with $\magg$ and $\stagg$ being non-conserved and conserved respectively. Further
\begin{align}
    \begin{split}
    \langle \xi^{\alpha}(x,t) \xi^{\beta}(x',t') \rangle &= A_M \delta^{\alpha \beta} \delta(x-x') \delta(t-t')
    \end{split}
    \label{eqn:Mnoise}
\end{align}
and
\begin{align}
    \begin{split}
\langle \zeta^{\alpha}(x,t) \zeta^{\beta}(x',t') \rangle &= - A_N \delta^{\alpha \beta} \nabla^2 \delta(x-x') \delta(t-t').
    \end{split}
    \label{eqn:Nnoise}
\end{align}

In the above equations, in addition to the relaxation time $\tau$, diffusion constant $D$ and noise $\boldsymbol{\xi}$ and $\boldsymbol{\zeta}$, we assume there are effective non-linear couplings $u_M$, $u_N$ and $u_{MN}$. The detailed derivation of the above equations by integrating out the fast modes is an involved exercise, which we defer to future work. Here, we motivate the form of the above equations from the structure of Eqns.(~\ref{eqn:M_contnus},~\ref{eqn:N_contnus}) with the observation that the parameters $u_M$, $u_N$ and $u_{MN}$, $\tau$, $D$ along with the noise strengths $A_M$ and $A_N$ are not independent and satisfy relations among themselves. These relations allow us to argue for the existence of diffusion of $\stagg$ and relaxation of $\magg$, while calculating the diffusion constant $D$ and relaxation time $\tau$ self-consistently. 

\begin{figure}[H]
    \centering
    \includegraphics[width=\linewidth]{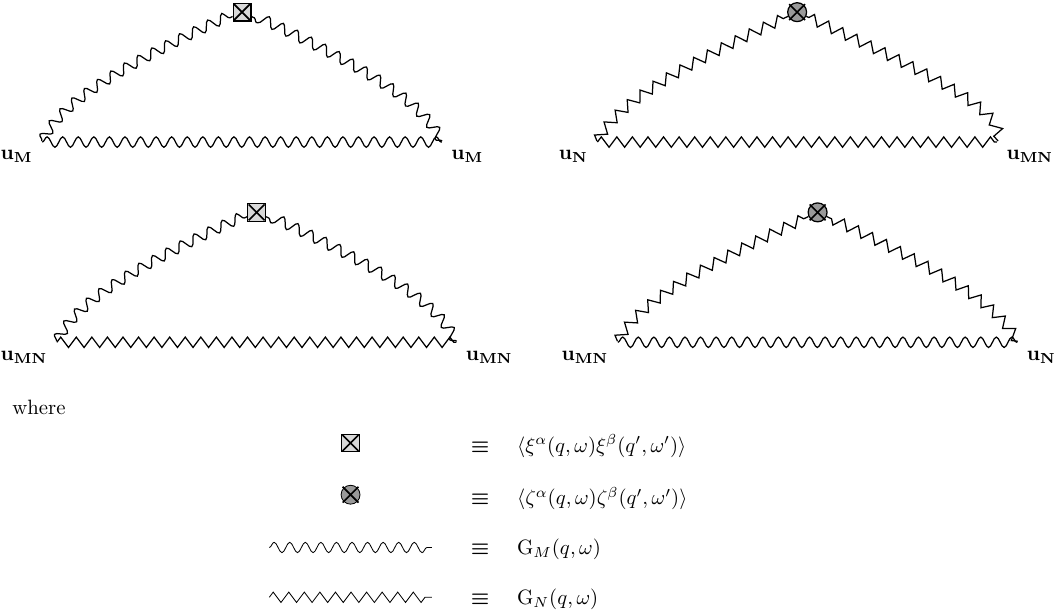}
    \caption{\small One-loop diagrams contributing to the self energy $\Sigma_M$ (top) and $\Sigma_N$ (bottom). $u_M$, $u_N$ and $u_{MN}$ are the coupling constants in Eqns.(~\ref{eqn:Mhydro},~\ref{eqn:Nhydro}). The momentum dependent vertex factors are also obtained from these equations. The gray shaded square is the correlator of the non-conserving noise(~\ref{eqn:Mnoise}), and the dark gray circle the correlator of the conserving noise(~\ref{eqn:Nnoise}). The two types of wavy lines are the propagators of the fields $\stagg$ and $\magg$.}
    \label{fig:Oneloop_FeynG}
\end{figure}

We proceed by adapting the formalism of~\cite{PhysRevB.11.4077} to our system. The Eqns.(~\ref{eqn:Mhydro},~\ref{eqn:Nhydro}) allow us to calculate the propagators for the fields $\magg$ and $\stagg$ perturbatively in the couplings $u_M$, $u_N$ and $u_{MN}$. The self-energy $\Sigma_M$ for the field $\magg$ is then set equal to $-1/\tau$ and the self-energy $\Sigma_N$ for $\stagg$ to $-Dq^2$ (where $q$ is the wavenumber) to obtain self-consistent equations for the relaxation time and the diffusion constant. Fig. (~\ref{fig:Oneloop_FeynG}) shows the diagrams that contribute to the self-energies upto one loop. 

These yield the following equations~\cite{Suppinf}
\begin{align}
    \begin{split}
\frac{1}{\tau} &= u_M^2A_M\tau^2\Lambda^3 + u_N u_{MN}A_N\Lambda/D^2 \\
D & = u_{MN}^2A_M\sqrt{\frac{\tau^3}{D}} + u_Nu_{MN}A_N \sqrt{\frac{\tau}{D^3}},
    \end{split}
    \label{eqn:selfcons}
\end{align}
where we have absorbed the numerical prefactors arising from the integrals in the calculation of the diagrams in Fig. (~\ref{fig:Oneloop_FeynG}) into the definitions of the effective parameters $u_M$, $u_N$, $u_{MN}$, $A_M$ and $A_N$. This allows us to obtain the terms on the right hand side of Eqn.(~\ref{eqn:selfcons}) simply from power counting of the diagrams in terms of the ultraviolet cutoff $\Lambda$ ~\cite{Suppinf}. Note that that the effective parameters in Eqns.(~\ref{eqn:Mhydro}) and(~\ref{eqn:Nhydro}) are, in principle, dependent on the values of $\stagg$ and ${\mathcal{E}}$ since these are the quantities conserved by the dynamics. 

To numerically demonstrate the presence of diffusion, we perform a calculation analogous to that performed for the magnetization and energy of the classical Heisenberg model in~\cite{das2018light}. We calculate the two point correlator $C_{N}(x,t)$ of $\stagg$ by sampling over all possible initial states with equal weight. This is an unbiased calculation over all microstates and is simpler to perform than one with fixed values of the conserved quantities. For the Heisenberg model, in the thermodynamic limit, it corresponds to a zero value of both the energy and the magnetization, from the equivalence of ensembles. For our model, it presumably corresponds to zero values of both $\stagg$ and ${\mathcal{E}}$ in the thermodynamic limit. 

\paragraph{Numerical results:} Each individual configuration of our system is  evolved according to  Eqn.(~\ref{eqn:drvn_dynmk}), with $\Delta t = 0.001 - 0.002 $ using an RK4 integrator. The system size for correlator calculation is $L= 512$, with  $5000$ initial conditions sampled from an ensemble in which all microstates are equally weighted. The correlator $C(x,t)$ is calculated by sliding over the reference point $x_0,t_0$, averaging over 500 consecutive snapshots for each fixed time t, for each initial state.
Under the ambit of hydrodynamic theory, the two-point correlator of conserved quantities shows a scaling function of the form $C(x,t) \sim t^{-\alpha} f(t^{-\alpha}x)$ , with $\alpha$ as the scaling exponent and $f$ a universal function. The numerical results for $C_{N}(x,t)$ and $C_{\mathcal{E}}(x,t)$ in Fig. (~\ref{fig:NN_EN_corrxt}) show that the two-point correlators for both the conserved quantities decay in time in a manner consistent with diffusion ($\alpha = 1/2$). The insets show a scaling collapse of both correlators to the diffusive form $t^{-1/2} f(t^{-1/2}x)$, where the exact function $f$ is different for the two cases. 

The magnetization starting from different initial states is found to rapidly decay to zero numerically as expected from relaxational dynamics for $\magg$~\cite{Suppinf}. In particular, for initial conditions in which the magnetization has a modulation at a wavevector $k$ about a uniformly magnetized state, the relaxation time is finite in the limit $k \rightarrow 0$ as expected due to the finite value of $\tau$ we have assumed in our theory. 
\begin{figure*}[htp]
  \centering
  \subfigure{\includegraphics[scale=0.4]{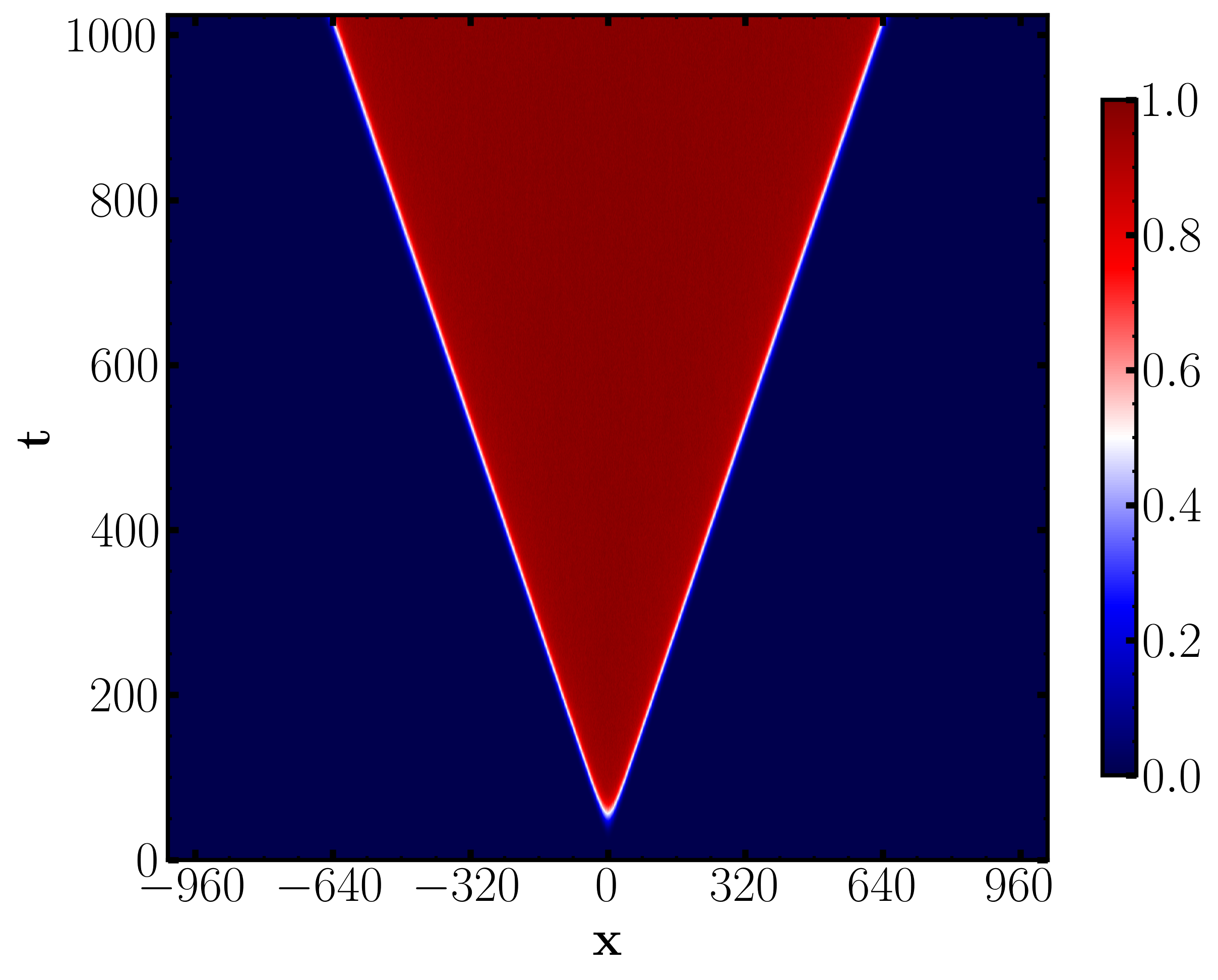}}
  \label{fig:puredrvn}
  \quad \quad
  \subfigure{\includegraphics[scale=0.41]{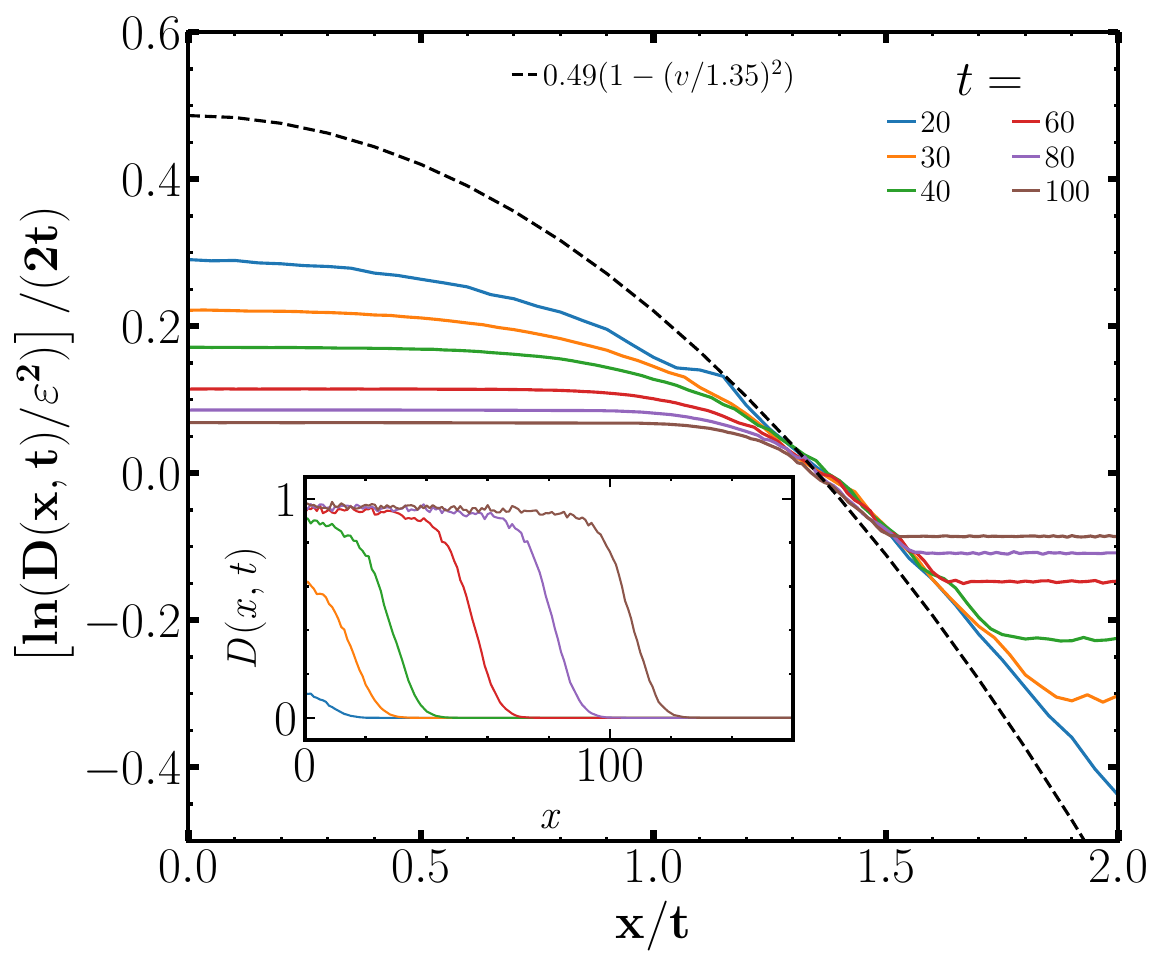}}
  \caption{\small (Left) Colormap of the decorrelator $D(x,t)$ calculated by averaging over pairs of initial conditions that differ only in the value of the spin ${\bf S}$ at the site $x=0$. It can be seen that this initial disturbance spreads ballistically from which the butterfly velocity $v_B=1.35$ can be obtained. (Right) The decorrelator given by the expression in Eqn.(~\ref{eqn:front}) plotted as a function of $x/t$ for different values of $t$. As expected, it can be seen that there is a collapse of the curves in the vicinity of the front from which the Lyapunov exponent $\approx 0.49$ can be extracted. The inset shows $D(x,t)$ as a function of $x$ for different values of $t$. The existence of a front can be seen from the rapid decrease in the value of $D(x,t)$ as a function of $t$ (inset).}
  \label{fig:Dxt_logDxt_drvn}
\end{figure*}

\paragraph{Decorrelator and chaos propagation:}
We now turn to a different characterization of the dynamics of our system, namely the spreading of chaos. Chaotic dynamics is considered to be an important indicator of the presence of thermalization in Hamiltonian systems. Chaos spreading in the classical Heisenberg chain has been quantified through a numerical calculation of the decorrelator, which shows a ballistic spreading of chaos with a butterfly velocity~\cite{das2018light}. Further, a calculation of the decorrelator allows one to obtain a Lyapunov exponent even for spin systems, whose elementary degrees of freedom do not have a symplectic Poisson bracket structure. We perform a similar calculation of the decorrelator for our system. The decorrelator is defined as
\begin{align}
    D( x_i, t) = 1- \langle \spin_i^{a}( t)\cdot \spin_i^{b}( t) \rangle,
    \label{eqn:Decorr}
\end{align} 
where $a$ and $b$ are two different initial conditions which differ only at a single site of the lattice ($i=0$) as 
\begin{align*}
\spin^{b}_0(0) = \spin^{a}_0(0) + \delta \spin_0
\end{align*} 
where
\begin{align}
\begin{split}
\delta \spin_0 &= \varepsilon[\hat{\mathbf{n}} \times \spin^a_0(0)] \\
\hat{\mathbf{n}} &= [\hat{\mathbf{z}} \times \spin^a_0(0)]/|\hat{\mathbf{z}} \times \spin^a_0(0)|
\end{split}
\label{eqn:Epsilonchange}
\end{align}
$\hat{\mathbf{z}}$ is the unit-vector along an arbitrarily define z-axis and $\varepsilon$ is the strenth of the perturbation (i.e. a measure of how different the two initial conditions are). As for the calculation of the two point correlators, all possible initial conditions are equally weighted while performing the average $\langle \dots \rangle$ in Eqn.(~\ref{eqn:Decorr}). 

We used $L=2048$, $\Delta t = 0.001$ in the numerical evolution of (~\ref{eqn:drvn_dynmk}), averaged over 10000 uniformly drawn initial configurations, to evaluate the decorrelator. The spin arrays were stored with quadruple numerical precision.  A colormap of the decorrelator obtained is shown in Fig.(~\ref{fig:Dxt_logDxt_drvn}). It can be clearly seen that the disturbance of strength $\varepsilon$ introduced at site $i=0$ spreads ballistically with a clearly defined butterfly velocity $v_B$. We obtain $v_B = 1.35(\pm 0.02) $ which is different from the value obtained for the classical Heisenberg model~\cite{das2018light} illustrating the difference in the microscopic dynamics of the two models.     

In the vicinity of the ballistically propagating front, the form of the decorrelator can be approximated as~\cite{das2018light} as
\begin{align}
\log
\dfrac{D(x,t)}{\varepsilon^2 } &= 2 \kappa t [1 - (x/v_B t)^2],
\label{eqn:front}
\end{align}
where $\kappa$ is the Lyapunov exponent of the system. A fit to this form in the vicinity of the front is also shown in Fig. (~\ref{fig:Dxt_logDxt_drvn}) from which one can obtain a Lyapunov exponent as well equal to $0.49(\pm 0.02)$. This value falls within the error range of the value calculated for the Heisenberg model~\cite{das2018light,Suppinf}.

\paragraph{Discussion:} To summarize, we have studied the dynamics of a classical spin chain with non-reciprocal couplings, which cannot be derived from a Hamiltonian employing standard Poisson bracket relations. We find that the dynamics conserves the staggered magnetization and a pseudo-energy which are analogous to the conserved magnetization and energy of the classical Heisenberg model. We have developed a hydrodynamic description of the model which self-consistently produces diffusion for the staggered magnetization and a relaxation of the magnetization for which we have also provided numerical evidence by simulating the microscopic dynamics. We have also shown that the model we study displays a ballistic spreading of chaos 
surprisingly similar to that seen in the Heisenberg model and we have numerically obtained the butterfly velocity and Lyapunov exponent. 

Our primary conclusion is that there is clearly an analogue of thermalization in the model we have studied despite the absence of a Hamiltonian. The emergence 
of diffusion in conserved quantities from an underlying precessional dynamics, and the ballistic spreading of chaos in strict analogy with the Heisenberg model --- for which these phenomena are presented as evidence of thermalization--- 
form the basis of our conclusion. 
Our findings open up the possibility of expanding the notion of thermalization 
to encompass the  behaviours observed in non-reciprocal systems such as the one we study. An investigation of the nature of the long-time steady state and a statistical description of it in terms of suitable ensembles is a fertile direction for future work.

\paragraph{Acknowledgements:} SR acknowledges the SERB, India, for support in the form of a J C Bose National Fellowship. SM acknowledges support from the DST, Govt. of India. NB acknowledges support from UGC India.

\newpage

\renewcommand{\theequation}{S.\arabic{equation}}
\setcounter{equation}{0}

\centerline{\large{\bf Supplemental information}}

\vspace*{0.5cm}

\subsection*{Derivation of the hydrodynamic equations}
The equations of motion for the odd and even spins are 
 \begin{align}
 \label{eqn:motion}
 \begin{split}    
 \dot{\spin}_{2i} = \spin_{2i} \times (\spin_{2i+1} - \spin_{2i-1}) \\
\dot{\spin}_{2i+1} = \spin_{2i+1} \times (\spin_{2i+2} - \spin_{2i})
\end{split}
\end{align}
We define the local magnetization $\magg_i$ and the local staggered magnetization $\stagg_i$ on the bond connecting spins $\spin_{2i}$ and $\spin_{2i+1}$ as
\begin{align}
\label{eqn:defMN}
\begin{split}
\magg_i & = \frac{\spin_{2i} + \spin_{2i+1}}{2} \\
\stagg_i & = \frac{\spin_{2i} - \spin_{2i+1}}{2}
\end{split}
\end{align}
Using eqns.(~\ref{eqn:motion}) and (~\ref{eqn:defMN}), we can write down the equations of motion for $\magg_i$ and $\stagg_i$

\begin{align}
    \label{eqn:contnutMN_discrete}
    \begin{split}
        \dot{\magg}_i &= \magg_i \times \Delta \magg_i - \stagg_i \times \Delta \stagg_i \\
        \dot{\stagg}_i &= \stagg_i \times \Delta \magg_i - \magg_i \times \Delta \stagg_i, 
    \end{split}
\end{align}
where 
\begin{align*}
\begin{split}
\Delta \magg_i &= \magg_i - \magg_{i-1} \\
\Delta \stagg_i &= \stagg_i - \stagg_{i-1}
\end{split}
\end{align*}

Taking the continuum limit of Eqn.(~\ref{eqn:contnutMN_discrete}) retaining only the first order terms, we obtain the hydrodynamic equations for $\magg$ and $\stagg$.
\begin{align}
\label{eqn:contnuteqn_puredrvn}
\begin{split}
\partial_t \magg &=  \magg \times \partial_x \magg  - \stagg \times \partial_x \stagg \\
\partial_t \stagg &= \partial_x (\stagg \times \magg) 
\end{split}
\end{align}
The local pseudo-energy is
\begin{align*}
\begin{split}
\Tilde{\mathcal{E}}_i &= \frac{1}{2}\left[\spin_{2i}.\left(\spin_{2i-1}- \spin_{2i+1}\right) \right]
\end{split}
\end{align*} 
from which we obtain 
\begin{align*}
\begin{split}
\partial_t \Tilde{\mathcal{E}}_i &= J^E_i - J^E_{i+1},
\end{split}
\end{align*}
where 
\begin{align*}
\begin{split}
J^E_i &= \frac{\spin_{2i-1}}{2}.\left(\spin_{2i} \times \spin_{2i-2}\right)
\end{split}
\end{align*}

Writing the pseudo-energy equations in terms of $\magg_i$ and $\stagg_i$ and taking the continuum limit,
\begin{align*}
    \Tilde{\mathcal{E}} &=-\frac{1}{2} \left(\magg + \stagg \right)\cdot \partial_x \left(\magg - \stagg \right) 
\end{align*}
and using the following equations based on  (~\ref{eqn:contnuteqn_puredrvn}),
\begin{align*}
    \partial_t (\magg + \stagg) = (\magg + \stagg) \times \partial_x (\magg - \stagg) \\
    \partial_t (\magg - \stagg) = (\magg - \stagg) \times \partial_x (\magg + \stagg)
\end{align*}
we obtain 
\begin{align}
\begin{split}
\partial_t \Tilde{\mathcal{E}} &= -\frac{1}{2} \partial_t(\magg + \stagg) \cdot \partial_x (\magg - \stagg) \\
 &- \frac{1}{2}(\magg + \stagg) \cdot \partial_x \left(\partial_t(\magg - \stagg) \right) \\
 &= -\frac{1}{2} (\magg + \stagg) \cdot \partial_x \left ((\magg - \stagg) \times \partial_x (\magg + \stagg) \right) \\
		&= -\frac{1}{2} \partial_x \left( (\magg + \stagg) \cdot \left((\magg - \stagg) \times \partial_x (\magg + \stagg)\right) \right) \\
  &= \partial_x \left( (\magg \times \stagg) \cdot \partial_x (\magg + \stagg) \right)
\end{split} 
\label{eqn:pseudo}
   \end{align}
where we observe in the last step that triple products of the form $\partial_x(\magg + \stagg) \cdot (\magg \times \magg - \stagg \times \stagg)$ will be zero. The final equation above gives us the Eqn. (4) of the main paper.

\subsection*{The evolution of the magnetization and staggered magnetization}
\begin{figure}[H]
\centering
\includegraphics[scale=0.34]{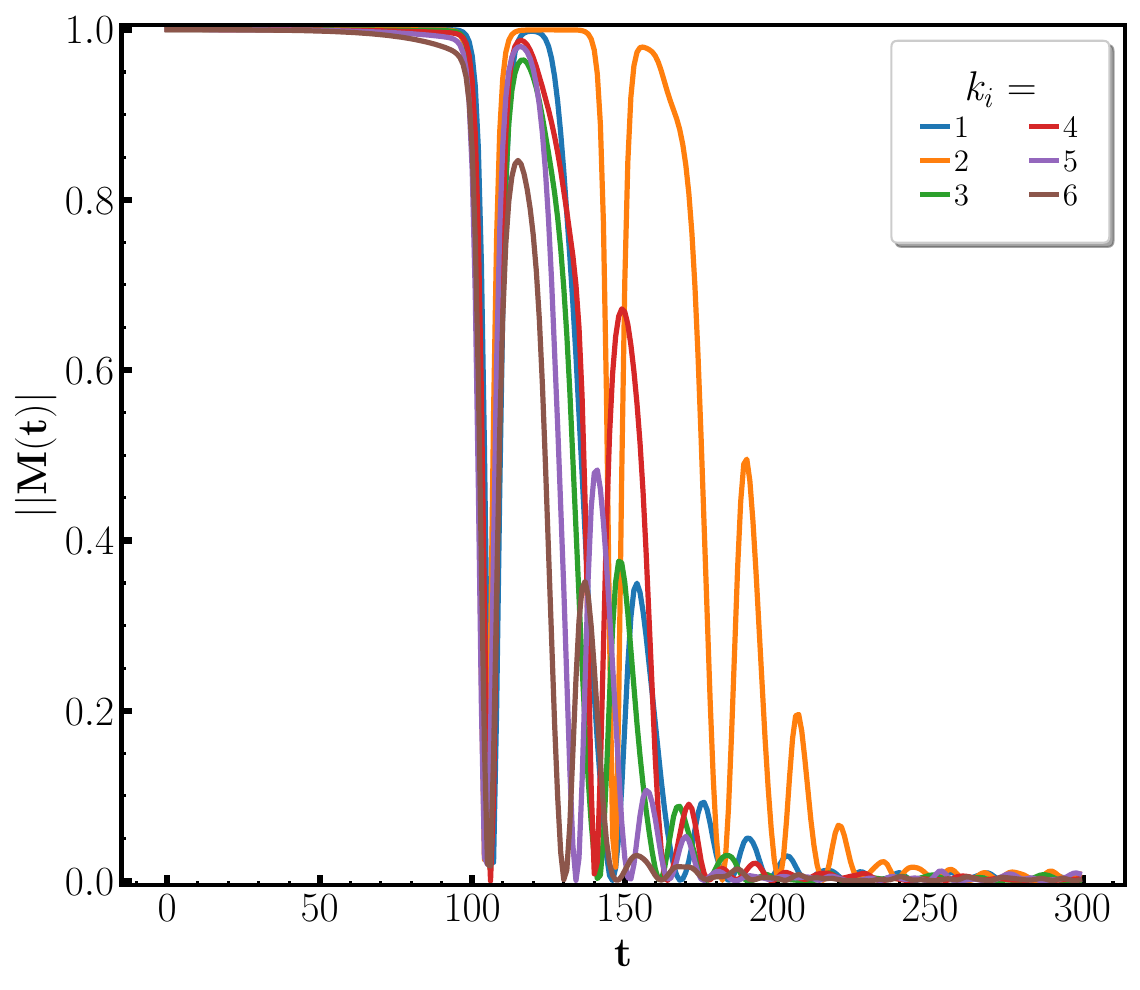}
\caption{\small The evolution of $|\magg(t)| = |\sum_i {\bf S}_i(t)|$ with time for different initial states of the form ${\bf S}_i(t=0) = \delta M \cos(2 \pi k i/L) \hat{e}_1 + \delta M \sin(2 \pi k i/L) \hat{e}_2 + M \hat{e}_3$. These represent states with a uniform value $M$ of one component of the spin with a modulation of strength $\delta M$ and wavevector $k$ of the other two components. The initial states thus have a total non-zero magnetization. It can be seen that this magnetization decays to zero as a function of time consistent with a non-zero decay time. The features (bounces) in the early time behavior are transients that cannot be captured by the hydrodynamic theory.}
\label{fig:Mkt_decayvs_t}
\end{figure}

Before we proceed, we show the diffusion behaviour in $\stagg$-dynamics can be obtained, to a first order approximation, from the steady state description of $\magg $ ($\partial_t \magg = 0$):
\begin{align}
    \begin{split}
    \magg & \approx -\tau (\stagg \times \partial_x \stagg) \\
    \partial_t \stagg &= \partial_x(\stagg \times \magg) + \mathbf{\zeta}\\
\Rightarrow \partial_t \stagg  &\approx -\tau \partial_x (\stagg \times (\stagg \times \partial_x \stagg)) + \boldsymbol{\zeta}  \\
    &= \tau \partial_x (\stagg^2(\mathbf{1} - \hat{\stagg}\hat{\stagg} \cdot ) \partial_x \stagg) + \boldsymbol{\zeta }
    \end{split}
    \label{eqn:Diffeqn_stagg}
\end{align}
The interesting behaviour of the magnetization is captured in the transient state. We evaluate the non-reciprocal dynamics for initial states with a non-zero magnetization: a small periodic modulation of the spins is added to a perfectly aligned state. It can be seen from Fig.(~\ref{fig:Mkt_decayvs_t}) that it decays fairly rapidly to zero for different values of the modulation wavenumber $k$. Further, the relaxation of the magnetization in the limit $k \rightarrow 0$ in a finite time is consistent with our assumption of a finite $\tau$ arising from the hydrodynamics. This can be seen in the form of a fairly rapid relaxation of the magnetization for the lowest value of $k$ (=1) possible in our numerics. A description of the structure (bounces) in the early time behavior of the magnetization is beyond the scope of the hydrodynamic framework.\\

\subsection*{Perturbation calculation for Self-energy terms}
We write the $\magg$-dynamics in the Fourier space, keeping the relaxation term but leaving out the non-linear contributions to zeroth order. 
\begin{align}
    \begin{split}
    (-i \omega  +\frac{1}{\tau}) M^{\alpha} &=  \xi^{\alpha} \\
    G_{M0}^{\alpha \beta} &= (-i \omega  +\frac{1}{\tau})^{-1} \delta^{\alpha \beta}
    \end{split}
    \label{eqn:M0propg}
\end{align}

The non-linear terms are introduced as perturbations, which contribute to the relaxation of $\magg$ through a self-energy term.
\begin{align}
    \begin{split}
    (-i \omega  +\frac{1}{\tau}) M(q, \omega)^{\alpha} &=  iq (u_M \epsilon^{\alpha \beta \gamma} M^{\beta} M^{\gamma} - u_N \epsilon^{\alpha \beta \gamma} N^{\beta} N^{\gamma})\\
    & +\xi^{\alpha}
    \end{split}
    \label{eqn:M1propg}
\end{align}

Similarly, writing the $\stagg$-dynamics in the Fourier space to the zeroth order
\begin{align}
    \begin{split}
    (-i \omega + Dq^2) N^{\alpha} &=  \zeta^{\alpha} \\
    G_{N0}^{\alpha \beta} &= (-i \omega  +Dq^2)^{-1} \delta^{\alpha \beta}
    \end{split}
    \label{eqn:N0propg}
\end{align}

and then adding the perturbation
\begin{align}
    (-i \omega  +Dq^2) N(k, \omega)^{\alpha} &=  iq (u_{MN} \epsilon^{\alpha \beta \gamma} N^{\beta} M^{\gamma} )+ \zeta^{\alpha}
    \label{eqn:N1propg}
\end{align}

We write the magnetization correlation function from the non-conserving noise
\begin{align}
    \label{eqn:magg_correlation_ft}
    \langle M^{\alpha} (q, \omega) M^{\beta} (q', \omega') \rangle =  \dfrac{4 \pi^2 A_M}{\omega^2 + \tau^{-2}} \delta^{\alpha \beta} \delta(q+q')  \delta (\omega + \omega')
\end{align}

The staggered magnetization correlation function is similarly:
\begin{align}
    \label{eqn:stagg_correlation_ft}
    \langle N^{\alpha} (q, \omega) N^{\beta} (q', \omega') \rangle =  \dfrac{4 \pi^2 A_N q^2}{\omega^2 + (Dq^2)^2} \delta^{\alpha \beta} \delta(q+q')  \delta (\omega + \omega')
\end{align}

Evaluating the propagator involves adding corrections from the higher resolution in the perturbation terms:
\begin{align*}
    G(q,\omega) = G_0 (q, \omega) + G_0 (q, \omega) \Sigma G (q, \omega)
\end{align*}
The propagators for $\magg$ and $\stagg$ are of the form $ \displaystyle G (q,\omega)= \frac{1}{-i\omega + \Sigma(q,\omega)}$, where $\Sigma(q,\omega)$ is the self energy.  The self energy for $\magg$, $\Sigma_M(q=0,\omega=0) = -1/\tau$ since $\magg$ is not a conserved field of the dynamics and is thus expected to relax. The self energy for $\stagg$, $\Sigma_N(q \rightarrow 0,\omega=0) = -Dq^2$ as $\stagg$ is conserved and thus expected to display diffusion. The structure of the propagator $G_{N0}(q, \omega) $ dictates that each power of $\omega$ goes with -2 powers of $q$. Similar correspondence must hold for $G_N(q, \omega) $ as well if the perturbation series is to converge in the $q, \omega \rightarrow 0$ limit.\\

\begin{figure}[H]
\centering
\begin{subfigure}
  \centering
  \includegraphics[width=\columnwidth]{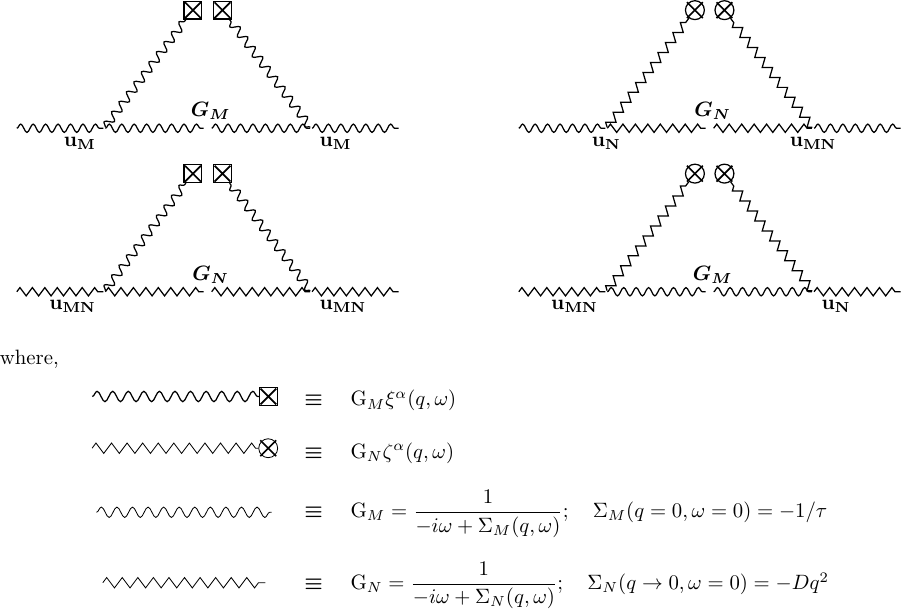}
\end{subfigure}
\caption{\small Vertex nodes for the self-energy terms for [\textit{Top}]: $\Sigma_M$ - incoming vertices are from $\magg$ , [\textit{Bottom}]: $\Sigma_N$ - incoming vertices are from $\stagg$; the circled cross depicts the conserved noise $\boldsymbol{\zeta}$ of strength $A_N$, and the open cross the non-conserved noise $\boldsymbol{\xi}$ with strength $A_M$.  }
\label{fig:oneloop_supp}
\end{figure}

We can obtain $\tau$ and $D$ in terms of the other parameters in eqns.(~\ref{eqn:M0propg}- ~\ref{eqn:N1propg}) self-consistently by calculating the above-mentioned self energies. This calculation to one loop is described below. The relevant diagrams are shown in Fig.(~\ref{fig:oneloop_supp}).  It can be seen that each diagram is given by sums of involving integrals of the form

\begin{align*}
    u_1 u_2 \int \int \dfrac{(\alpha_1 q + \beta_1 p)(\alpha_2 q + \beta_2 p)f(p)}{(\nu^2 + g(p)^2)(-i \nu + h(p))}~d\nu dp
\end{align*}
where $f(p)= \{A_M, A_N p^2 \}$, $g(p)= \{(Dp^2)^2,  \tau^{-1}\}$, $h(p)= \{Dp^2 , \tau^{-1}\}$, $u_1, u_2 = \{u_M, u_N, u_{MN} \}$. $\alpha_1$, $\beta_1$, $\alpha_2$ and $\beta_2$ are numerical constants and $p$ and $\nu$, the internal wavenumber and frequency. The factors in the denominator of the integrand come from the propagators, the factor $f(p)$ from the noise correlator and the other two factors of the form $\alpha q + \beta p$ come from the two vertices in each diagram since each is linear in the momenta of the associated propagators. The different integrals that contribute to a particular diagram are obtained from the different ways of dividing the external wavenumber $q$ and frequency $\omega$ into the wavenumbers $p$ and $q-p$ and frequencies $\nu$ and $\omega-\nu$ of the propagators in the loop in addition to choosing the different components of the vector fields $\magg$ and $\stagg$ in the loop. We are also only interested only in the leading order dependence in the ultraviolet cutoffs and so have replaced the propagators in the loop with their low wavenumber and low frequency forms and have taken the external frequency $\omega$ to be zero. As mentioned in the main text, numerical pre-factors from the above integrals can be absorbed into the definitions of the coupling strengths $u_M$, $u_N$ and $u_{MN}$ and the noise strengths $A_M$ and $A_N$. Thus, the contribution of each diagram can be obtained from a simple power counting of the contributing integrals. Performing the relevant integrals over the frequencies $\nu$ for each diagram and adding, we obtain 

\begin{align}
\label{eqn:loop_int}
   \Sigma (q,\omega=0) \sim - u_1 u_2 \int_{-\Lambda}^\Lambda \dfrac{(Eq^2 + Fp^2)f(p)}{g(p) \left[g(p)+h(p)\right]}dp,
\end{align}
for each of the diagrams in Fig.(~\ref{fig:oneloop_supp}), where $\Lambda$ is the ultraviolet cutoff for the wavenumber and $E$ and $F$ are numerical constants. Since, $f(p)$, $g(p)$ and $h(p)$ are all even functions of $p$, there are no terms linear in $p$ in the numerator of the integrand. 

The diagrams in the top row of Fig.(~\ref{fig:oneloop_supp}), contribute to $\Sigma_M$. For these, we set $q=0$ in Eqn.(~\ref{eqn:loop_int}). For the diagram on the left, $u_1=u_2=u_M$, $f(p)=A_M$, $g(p)=h(p)=\tau^{-1}$, producing
a contribution
\begin{align*}
   -u_M^2 A_M \tau^2 \int_{-\Lambda}^\Lambda p^2 dp \sim -u_M^2 A_M \tau^2 \Lambda^3
\end{align*}
For the diagram on the right, $u_1=u_N$, $u_2=u_{MN}$ $f(p)=A_Np^2$, $g(p)=Dp^2$,$h(p)=\tau^{-1}$, which gives
\begin{align*}
   -u_N u_{MN} \frac{A_N}{D} \int_{-\Lambda}^\Lambda \frac{p^2} {Dp^2 + \tau^{-1}}dp \sim -u_N u_{MN} A_N \frac{\Lambda}{D^2}
\end{align*}
Thus, we obtain the first of the two self-consistent equations 
\begin{align*}
\frac{1}{\tau} &= u_M^2A_M\tau^2\Lambda^3 + u_N u_{MN}A_N\Lambda/D^2
\end{align*}

The diagrams in the bottom row of Fig.(~\ref{fig:oneloop_supp}), contribute to $\Sigma_N$. For these, it turns out that $F=0$ and so they give contributions $\sim q^2$ as expected from the diffusive behavior of $\stagg$. For the diagram on the left, $u_1=u_2=u_{MN}$, $f(p)=A_M$, $g(p)=\tau^{-1}$ and $h(p)=Dp^2$, yielding
\begin{align*}
   -q^2u_{MN}^2 A_M \tau \int_{-\Lambda}^\Lambda \frac{1}{Dp^2+\tau^{-1}} dp \sim -q^2u_{MN}^2A_M\sqrt{\frac{\tau^3}{D}}
\end{align*}
Finally, for the diagram on the right, $u_1=u_{MN}$, $u_2=u_N$ $f(p)=A_Np^2$,  $g(p)=Dp^2$, $h(p)=\tau^{-1}$, giving
\begin{align*}
-q^2u_{MN}u_N \frac{A_N}{D} \int_{-\Lambda}^\Lambda \frac{1} {Dp^2 + \tau^{-1}} dp \sim -q^2u_Nu_{MN}A_N \sqrt{\frac{\tau}{D^3}}
\end{align*}
From these we obtain the other self-consistent equation
\begin{align*}
D & = u_{MN}^2A_M\sqrt{\frac{\tau^3}{D}} + u_Nu_{MN}A_N \sqrt{\frac{\tau}{D^3}}.
\end{align*}

\begin{figure*}[htp]
\centering
\subfigure{\includegraphics[scale=0.28]{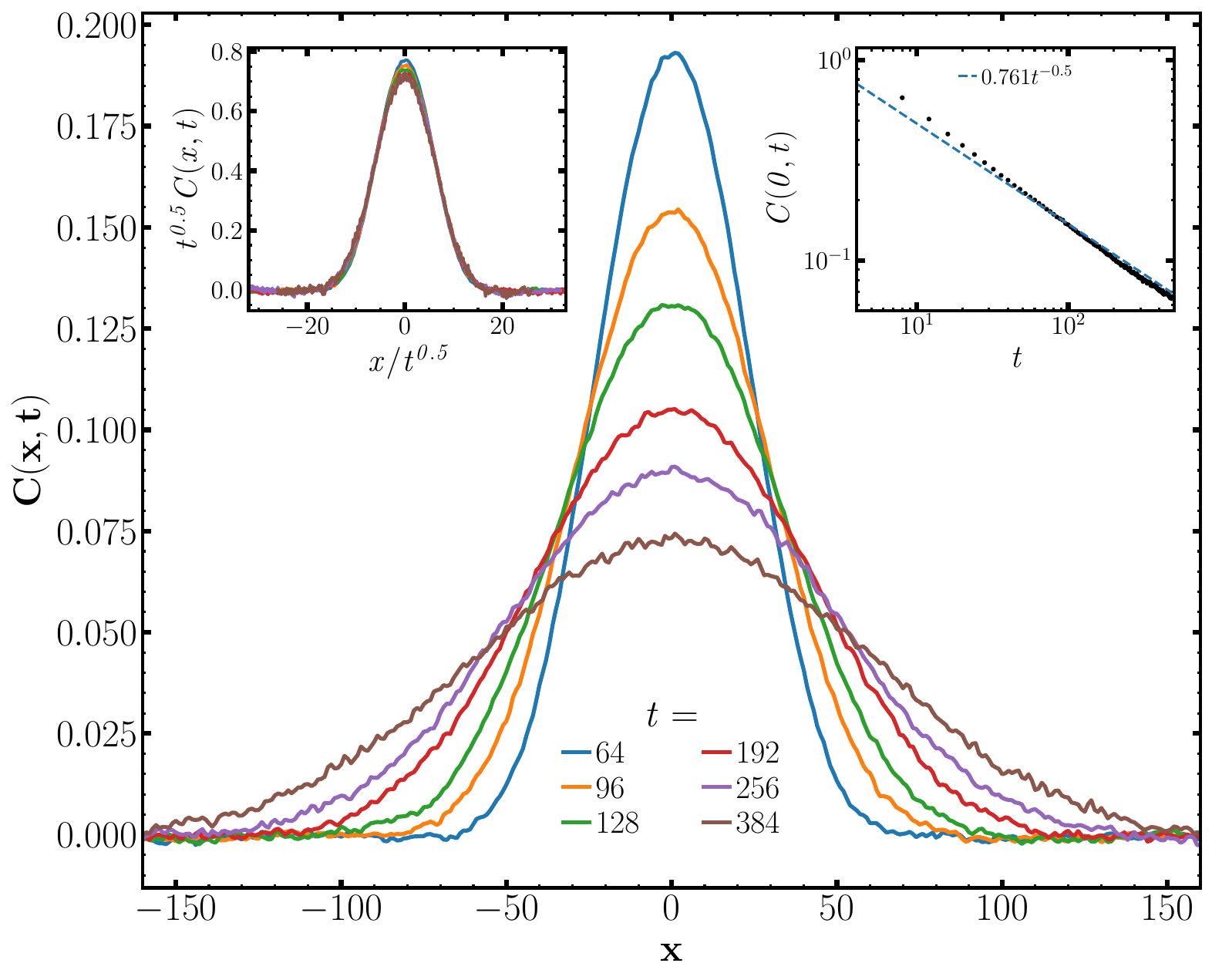}}
\quad \quad 
\subfigure{\includegraphics[scale=0.28]{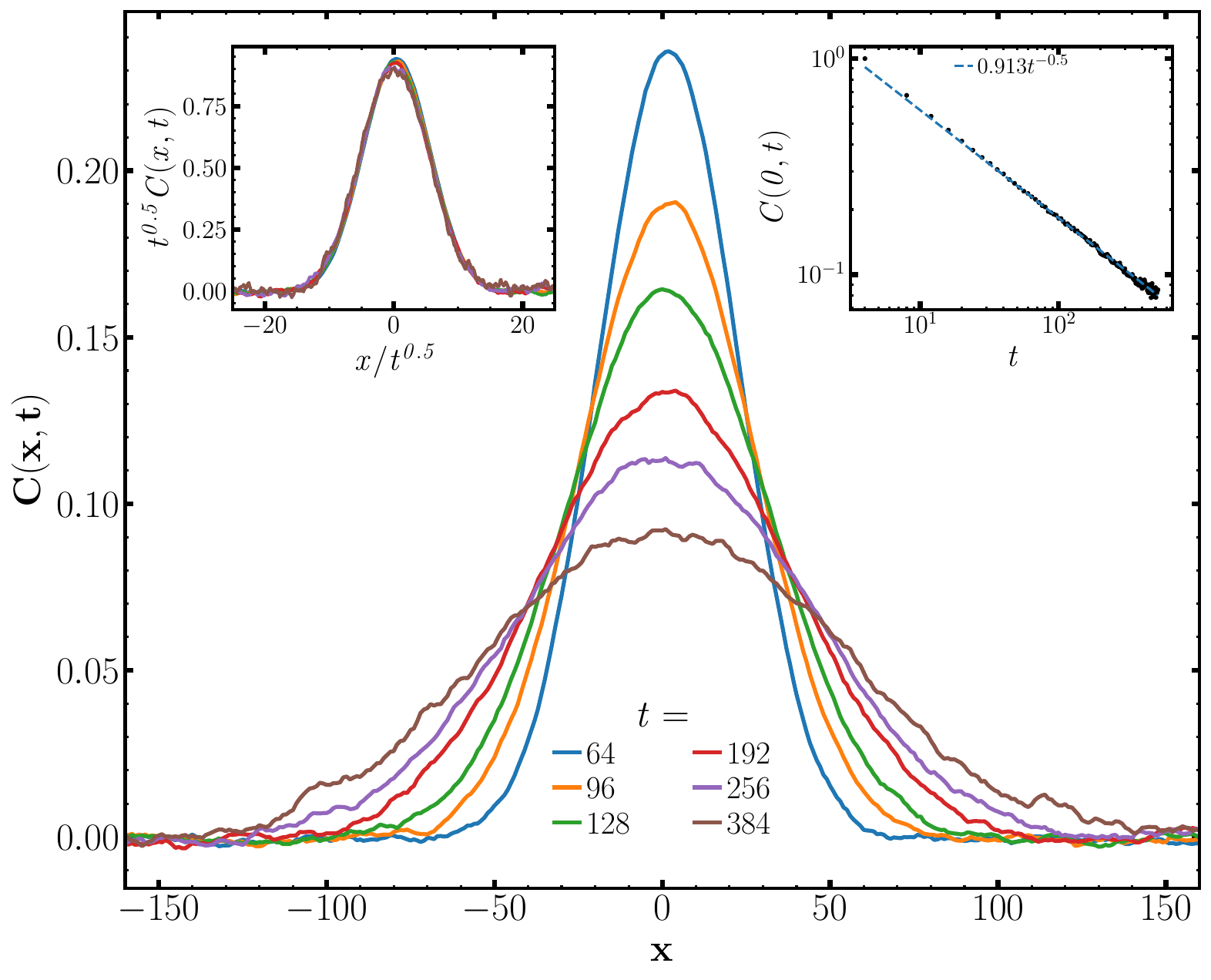}}
\caption{\small Two-point correlation functions of: (a) the standard magnetization, $C_{M}(x,t)$ and (b) the energy , $C_{E}(x,t)$ evaluated for the Heisenberg model. $x \in \{1,L\}$ for $L=512$, and over 5000 initial configurations sampled for averaging.  The left inset of both panels shows the scaling collapse to a form $C(x,t) = t^{-1/2}f(x/t^{1/2})$ consistent with diffusion while the right shows a plot of $C(0,t)$ versus $t$ with a fit to $t^{-1/2}$.}
\label{fig:MM_E_corrxt}
\end{figure*}


\begin{figure*}[htp]
\centering
\subfigure{\includegraphics[scale=0.4]{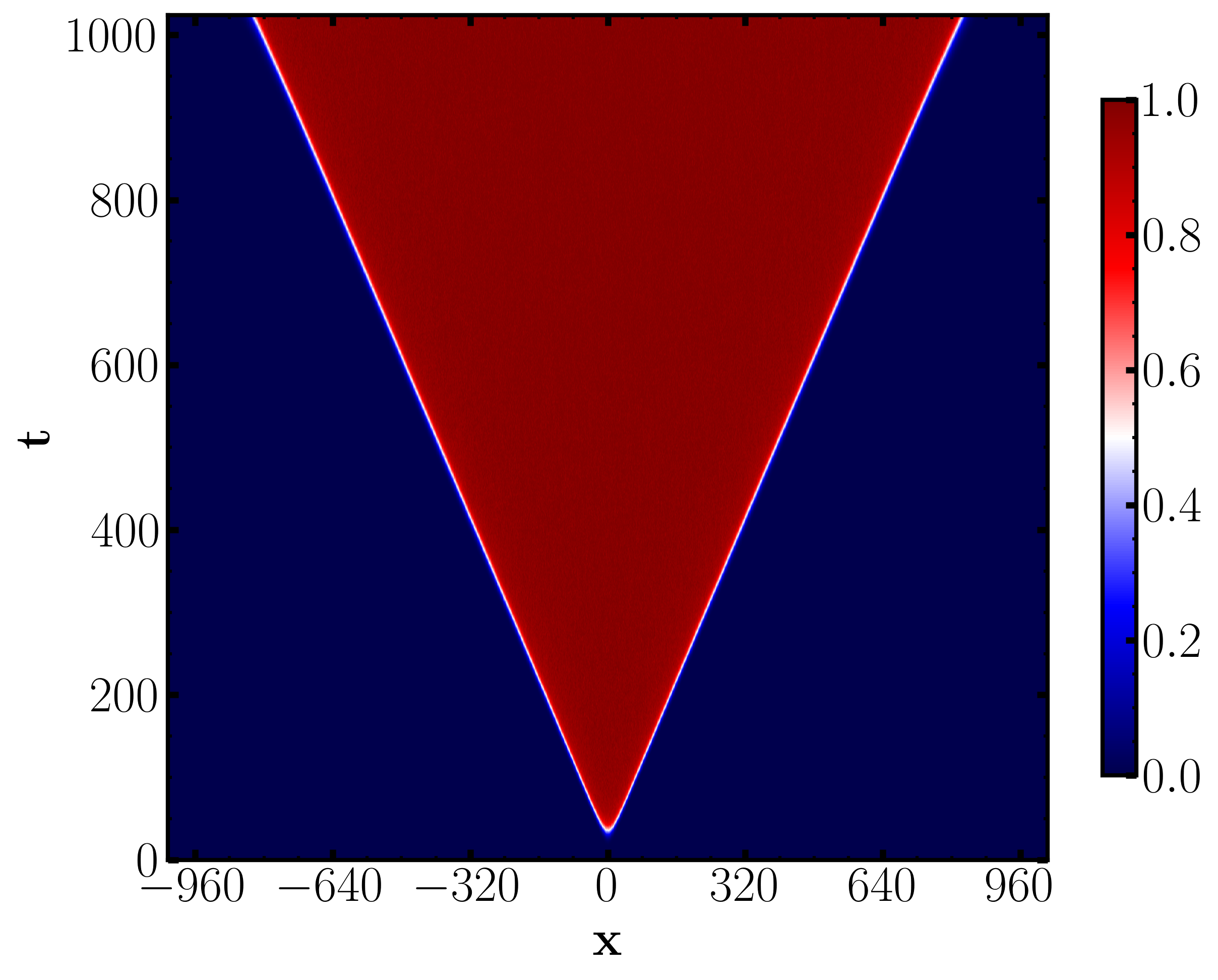}}
\quad \quad
\subfigure{\includegraphics[scale=0.41]{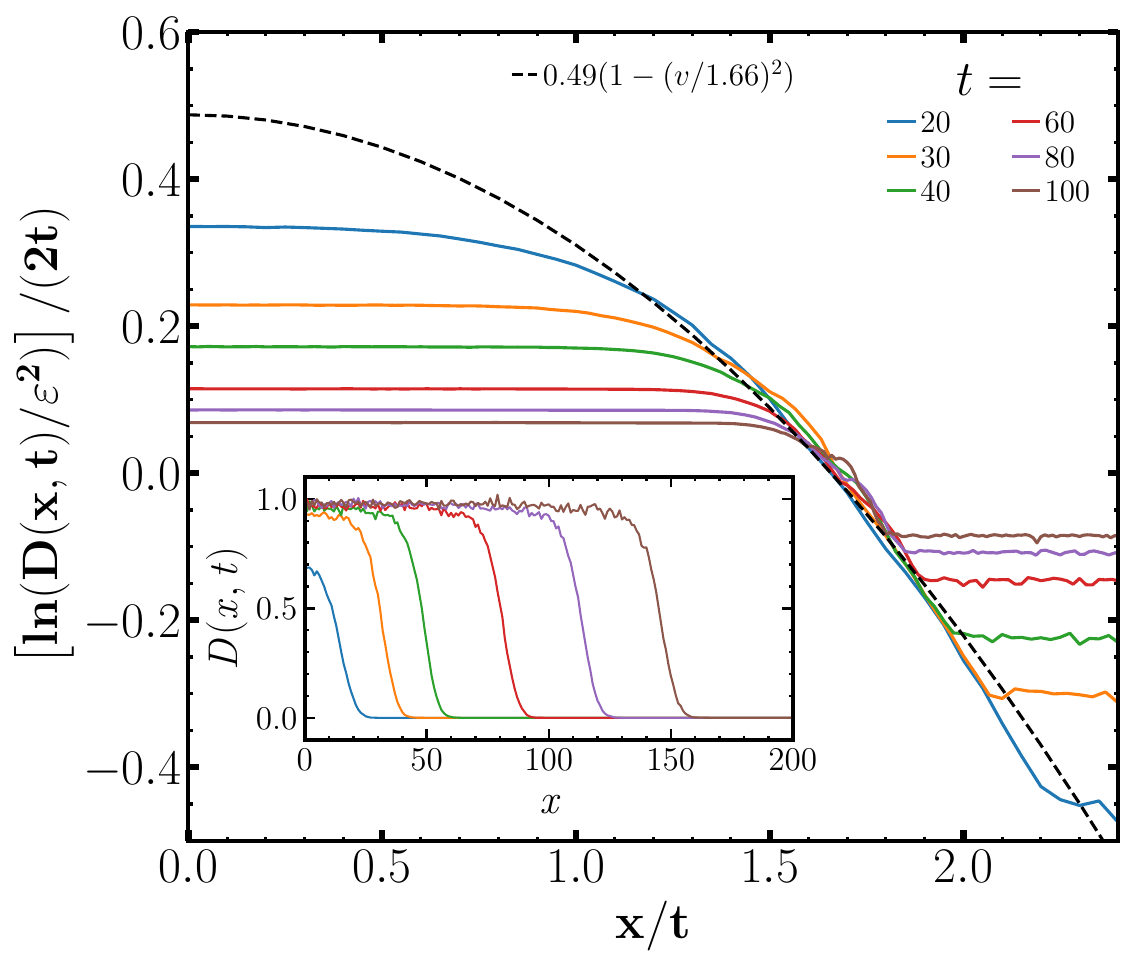}}
 \caption{\small \textit{(Left)} Colormap of the decorrelator $D(x,t)$ calculated by averaging over pairs of initial conditions that differ only in the value of the spin ${\bf S}$ at the site $x=0$. It can be seen that this initial disturbance spreads ballistically from which the butterfly velocity $v_B=1.66$ can be obtained. \textit{(Right)} The decorrelator given by the expression $\log(D(x,t)/\varepsilon^2) = 2 \kappa t (1 - (x/v_Bt)^2)$ plotted as a function of $x/t$ for different values of $t$. As expected, it can be seen that there is a collapse of the curves in the vicinity of the front from which the Lyapunov exponent $\approx 0.49$ can be extracted. The inset shows $D(x,t)$ as a function of $x$ for different values of $t$. The existence of a front can be seen from the rapid decrease in the value of $D(x,t)$ as a function of $t$ (inset).}
  \label{fig:Dxt_logDxt_hsbg}
\end{figure*}

\section*{Comparison with the Heisenberg dynamics}
To test the accuracy of our numerical code, we first ran the simulation for the Hamiltonian originating Heisenberg dynamics and obtained results for the conserved quantities, namely the standard magnetization and energy density. We confirmed that the conserved quantities show diffusive behaviour (Fig.~\ref{fig:MM_E_corrxt}) , thus setting up a standard to corroborate our results with. We also calculated the numerical values for the butterfly velocity and Lyapunov exponent for this case, and found them to be within expected error range of our simulation parameters, $v_B = 1.66(\pm 0.02) , \kappa = 0.49 (\pm 0.02)$ (Fig. ~\ref{fig:Dxt_logDxt_hsbg}), with $\varepsilon = 10^{-3}, L = 2048, \Delta t = 0.001-0.002$.  Finding $\kappa$ from the expression $\log(D(x,t)/\varepsilon^2) = 2 \kappa t (1 - (x/v_Bt)^2)$  requires knowledge of $D(0,t)$ but the perturbation strength $(10^{-3})$ is not small enough for the numerics to find an appreciable jump in $\log(D(x,t)/\varepsilon^2)$ from $t=0$ itself, as evident from the inset of the plot.\\ 

\begin{figure*}[htp]
\centering
  \subfigure{\includegraphics[scale=0.28]{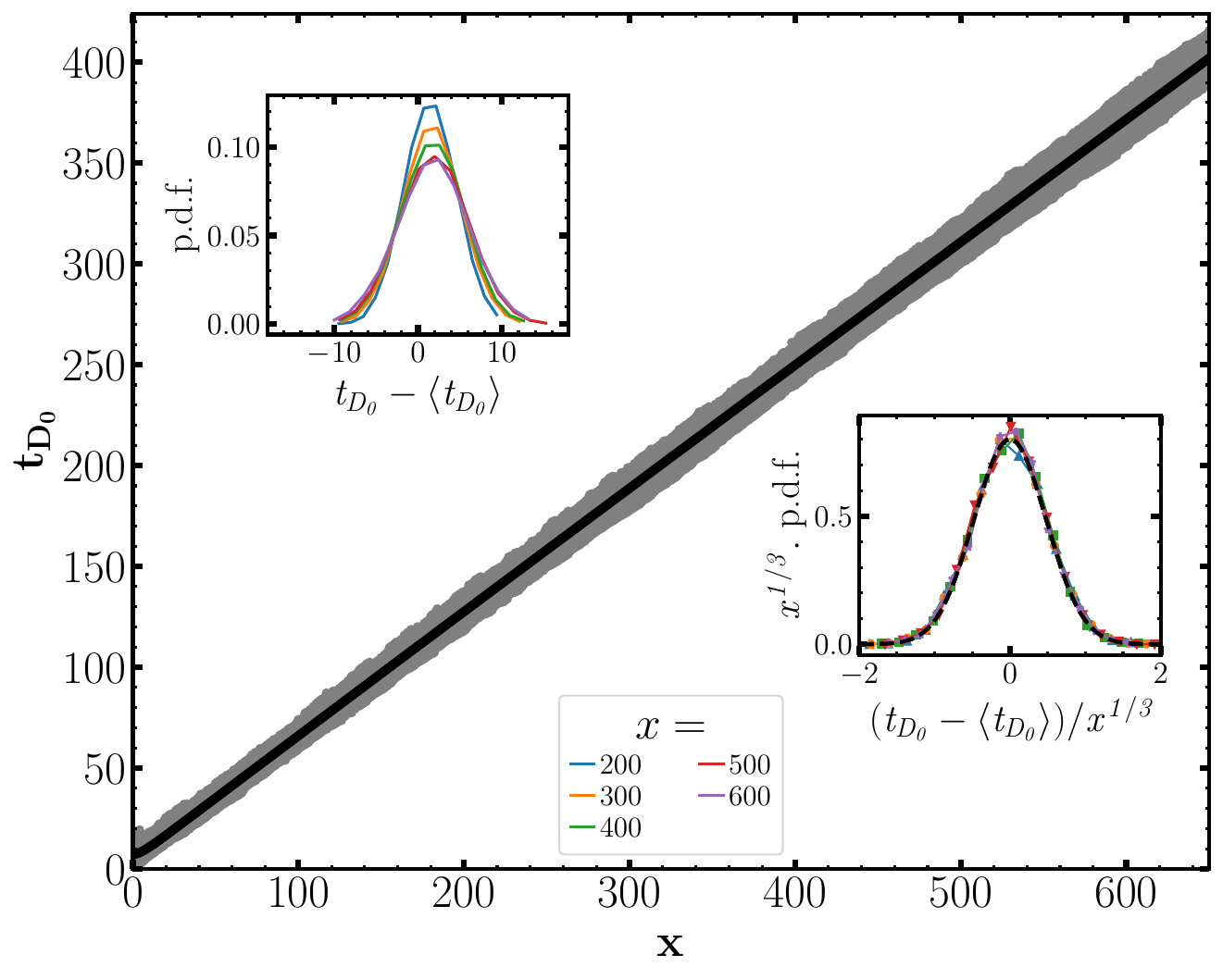}}           \quad \quad  \quad
  \subfigure{\includegraphics[scale=0.28]{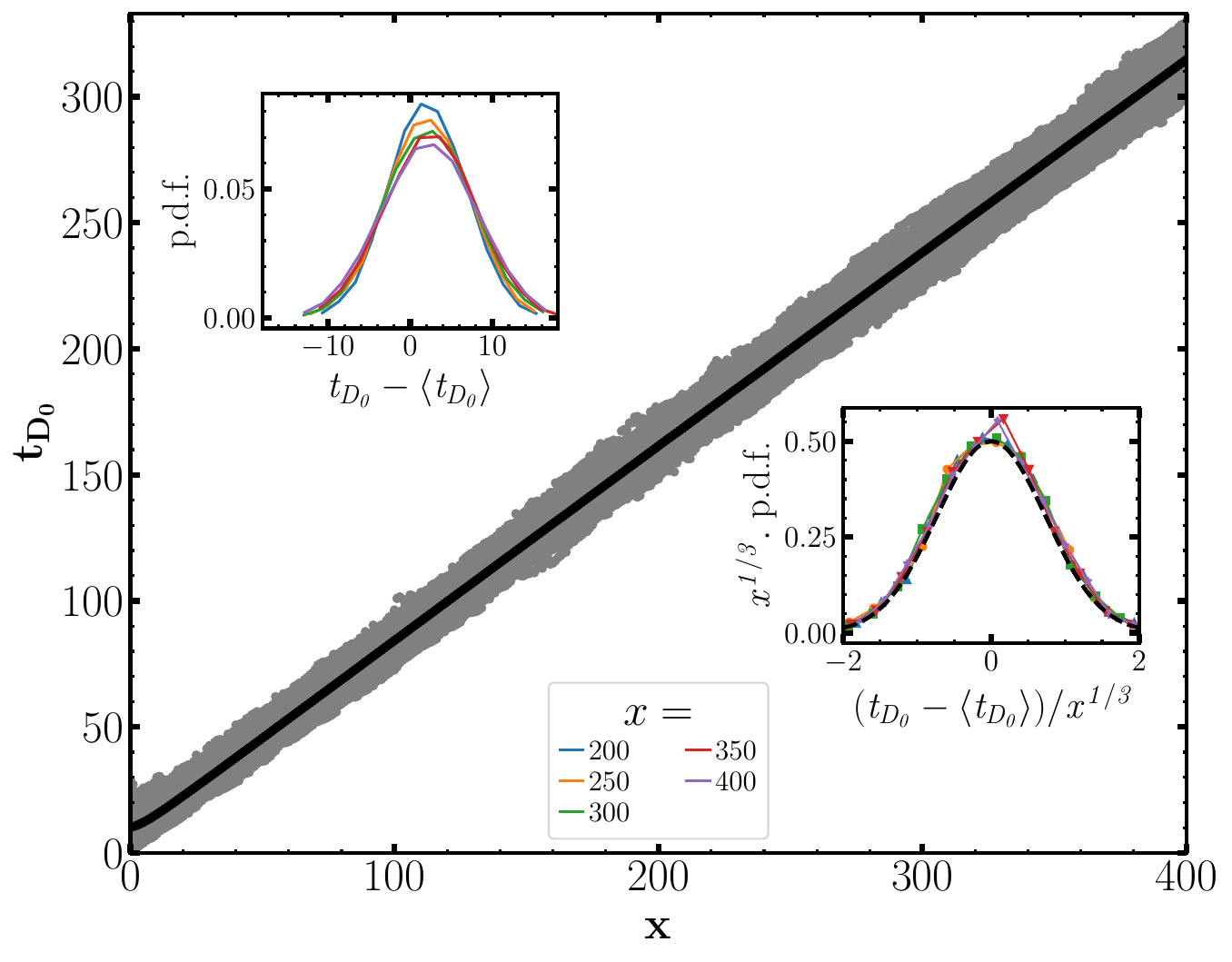}}
\caption{\small \textit{(Left, Right:)} Arrival-time plots for the Heisenberg and our driven model respectively. \textit{(Main panel:)}  Arrival times $t_{D_0}$ for the decorrelator front at a given site $x$, for $D_{0} = 100 \varepsilon^2 = 10^{-4}$. The central black line is the average of such arrival times calculated for individual configurations (grey scatter plot), whose slope gives us $v_B \approx 1.66, 1.35$ respectively. (\textit{Upper inset:}) The distributions of arrival times centered at the mean show diminishing peak and broadening variance with higher values of $x$. \textit{(Lower inset:)} The probability distribution functions (p.d.f.s) collapse when scaled with respect to site as $x^{1/3}$. A gaussian fit to the fluctuations is plotted for each case, with $\langle t_{D_0} \rangle$ centered at $x=600, 350$ for the Heisenberg, driven models respectively.}
\label{fig:Arrivaltime_sup}
\end{figure*}

Finally, we also find the power-law associated with the broadening of the decorrelator front (Fig.~\ref{fig:Arrivaltime_sup}). The approach here involves comparing the Decorrelator to a threshold value, $D_{0} = 100 \varepsilon^2$ and mark the time at which the single configuration arrival-front exceeds this threshold value, $D(x,t) \geq D_{0}$. 
Collecting this data for several samples ($\sim 10^4$) we see that the arrival-front of the decorrelator broadens with time. The distribution of the arrival-times from the mean arrival-time (the slope of which with respect to the position is just the inverse arrival velocity), shows a collapse when fit to a $1/3$ power-law.

\subsection*{Liouville's Theorem holds for the generalized nearest-neighbour precessional dynamics }

The canonical dynamics of SO(3) spins on a lattice is governed by
\begin{align}
    \label{eqn:canonical_dyn}
    \dfrac{d S_{i}^{\alpha}}{dt} &= \sum_{\beta \gamma}\epsilon_{\alpha \beta \gamma}S_i^{\beta}\dfrac{\partial H}{\partial S_i^{\gamma}}
\end{align}

so that
\begin{align}
    \sum_{\alpha} \dfrac{\partial \dot{S}_i^{\alpha}}{\partial S_i^{\alpha}} = \sum_{\alpha \beta \gamma} \epsilon_{\alpha \beta \gamma}\left(\dfrac{\partial S_i^{\beta}}{\partial S_i^{\alpha}}\dfrac{\partial H}{\partial S_j^{\gamma}} + S_i^{\beta}\dfrac{\partial^2 H}{\partial S_i^{\alpha} \partial S_i^{\gamma}}\right)
\end{align}
The terms in the parenthesis being symmetric in $\alpha \beta$ and $\alpha \gamma$ are eliminated by $\epsilon_{\alpha \beta \gamma}$. Thus the velocity in $S$-space is divergence free.

This argument holds true even for the generalized dynamics:
\begin{align*}
\dot{S}_{i \alpha} = \epsilon_{\abc}S_{i \beta}(S_{i+1 \gamma} \pm S_{i-1 \gamma})    
\end{align*}

\begin{align}
    \sum_{\alpha} \dfrac{\partial \dot{S}_i^{\alpha}}{\partial S_i^{\alpha}} = \sum_{\abc} \epsilon_{\abc}\left(\dfrac{\partial S_i^{\beta}}{\partial S_i^{\alpha}} (S_{i+1 \gamma} \pm S_{i-1 \gamma}) \right)
\end{align}
The above term vanishes since $\dfrac{\partial S _{j \alpha}}{\partial S_{i \beta}} = 0 $ for any $ \alpha, \beta, i \neq j$. \cite{2208.08577} presented a similar argument in their work on non-reciprocal spin models, starting with the Landau-Lifshitz equation.

\end{document}